\documentclass[12pt]{article}
\usepackage{cite}
\newcommand{\ba}{\begin{array}}
\newcommand{\ea}{\end{array}}
\newcommand{\be}{\begin{equation}}
\newcommand{\ee}{\end{equation}}
\newcommand{\bea}{\begin{eqnarray}}
\newcommand{\eea}{\end{eqnarray}}

\newcommand{\la}{\langle}
\newcommand{\ra}{\rangle}

\newcommand{\p}{\partial}

\def\CP{{\mathcal{P}}}
\def\CQ{{\mathcal{Q}}}
\def\IB{\relax\hbox{$\inbar\kern-.3em{\rm B}$}}
\def\IC{\relax\hbox{$\inbar\kern-.3em{\rm C}$}}
\def\ID{\relax\hbox{$\inbar\kern-.3em{\rm D}$}}
\def\IE{\relax\hbox{$\inbar\kern-.3em{\rm E}$}}
\def\IF{\relax\hbox{$\inbar\kern-.3em{\rm F}$}}
\def\IG{\relax\hbox{$\inbar\kern-.3em{\rm G}$}}
\def\IGa{\relax\hbox{${\rm I}\kern-.18em\Gamma$}}
\def\IH{\relax{\rm I\kern-.18em H}}
\def\IK{\relax{\rm I\kern-.18em K}}
\def\IL{\relax{\rm I\kern-.18em L}}
\def\IP{\relax{\rm I\kern-.18em P}}
\def\IR{\relax{\rm I\kern-.18em R}}
\def\IZ{\relax{\rm Z\kern-.5em Z}}

\def\half{\frac{1}{2}}
\def\p{\partial}
\def\f{\frac}

\begin{document}

\begin{titlepage}


\begin{flushright}
OUTP-02-11-P \\
December 2001\\
hep-th/0112094\\
\end{flushright}

\vskip 2 cm

\begin{center}
{\LARGE Extended chiral algebras in the $SU(2)_0$ WZNW model}
\vskip 1 cm

{\large A. Nichols\footnote{a.nichols1@physics.ox.ac.uk}}

\begin{center}
{\em  Theoretical Physics, Department of Physics, Oxford University\\
1 Keble Road, Oxford, OX1 3NP, UK } \\
\end{center}

\vskip 1 cm

\vskip .5 cm 

\begin{abstract}

We investigate the W-algebras generated by the integer dimension chiral primary operators of the $SU(2)_0$ WZNW model. These have a form almost identical to that found in the $c=-2$ model but have, in addition, an extended affine Lie algebraic structure. Moreover on Hamiltonian reduction these $SU(2)_0$ $W$-algebras exactly reduce to those found in $c=-2$. We explicitly find the free field representations for the chiral $j=2$ and $j=3$ operators which have respectively a fermionic doublet and bosonic triplet nature. The correlation functions of these operators accounts for the rational solutions of the Knizhnik-Zamolodchikov equation that we find. We explicitly compute the full algebra of the $j=2$ operators and find that the associativity of the algebra is only guaranteed if certain null vectors decouple from the theory. We conjecture that these algebras may produce a quasi-rational conformal field theory.

\end{abstract}

\end{center}

\end{titlepage}

\section{Introduction}

The study of conformal invariance in two dimensions has been a fascinating and productive area of research for the last twenty years \cite{Belavin:1984vu}. There is an interesting class of conformal field theories (CFTs) called logarithmic conformal field theories (LCFTs). In these theories the irreducible primary operators do not close under fusion and indecomposable representations are inevitably generated \cite{Gurarie:1993xq}. The operators in the theory have scaling dimensions that are either degenerate or differ by integers. In these cases it is possible to have a non-trivial Jordan block structure. 

LCFTs have emerged in many different areas for example: WZNW models and gravitational dressing \cite{Bilal:1994nx,Caux:1997kq,Kogan:1997nd,Kogan:1997cm,Giribet:2001qq,Gaberdiel:2001ny,Nichols:2001du,Kogan:2001nj}, polymers \cite{Saleur:1992hk,Cardy,Gurarie:1999yx}, disordered systems and the Quantum Hall effect \cite{Caux:1996nm,Kogan:1996wk,Maassarani:1996jn,Gurarie:1997dw,CTT,Caux:1998eu,Bhaseen:1999nm,Kogan:1999hz,Gurarie:1999bp,RezaRahimiTabar:2000qr,Bernard:2000vc,Bhaseen:2000bm,Bhaseen:2000mi,Ludwig:2000em,CardyTalk,Kogan:2002ku}, string theory 
\cite{Kogan:1996df,Kogan:1996zv,Ellis:1998bv,Ghezelbash:1998rj,Kogan:1999bn,Myung:1999nd,Lewis:1999qv,Nichols:2000mk,Kogan:2000nw,Moghimi-Araghi:2001fg},
 2d turbulence \cite{RahimiTabar:1996dh,RahimiTabar:1997nc,Flohr:1996ik,RahimiTabar:1997ki,RahimiTabar:1996si}, multi-colour QCD at low-x \cite{Korchemsky:2001nx} and the Seiberg-Witten solution of ${\mathcal N}=2$ SUSY Yang-Mills \cite{Cappelli:1997qf,Flohr:1998ew}. Deformed LCFTs, Renormalisation group flows and the $c$-theorem were discussed in  \cite{Caux:1996nm,Rahimi-Tabar:1998ph,Mavromatos:1998sa}. The holographic relation between logarithmic operators and vacuum instability was considered in \cite{Kogan:1998xm,Lewis:1998fg}

There has also been much work on analysing the general structure and consistency of such models in particular the $c_{p,q}$ models and the special case of $c=-2$ which is by far the best understood 
\cite{Kausch:1995py,Gaberdiel:1998ps,Kausch:2000fu,Flohr:2000mc}. It is unclear as yet how much of the structure, for instance the role of extended algebras, is generic to all LCFTs. For more about the general structure of LCFT see \cite{RahimiTabar:1997ub,Rohsiepe:1996qj,Kogan:1997fd,Flohr:2001tj} and references therein. Introductory lecture notes on LCFT and more references can be found in \cite{Tabar:2001et,Flohr:2001zs,Gaberdiel:2001tr,Moghimi-Araghi:2002gk}.

The WZNW model is of great importance in CFT. Correlation functions in such models obey differential, Knizhnik-Zamolodchikov (KZ), equations \cite{Knizhnik:1984nr} coming from null states in the theory. The solutions to the KZ equations  and correlation functions for the integrable sector of the $SU(2)_k$ model were studied by \cite{Zamolodchikov:1986bd,Christe:1987cy}. There is a simple Dotsenko-Fateev integral representation for solutions but these do not converge in many cases beyond the integrable representations. In particular in the cases in which logarithms appear we have to be very careful when analytically continuing the solutions. 

The $SU(2)_0$ WZNW model was studied in \cite{Kogan:1996df} in the context of describing NS-5 branes. Later $SU(2)_0$, and its supersymmetric extension, were studied in more detail \cite{Caux:1997kq}. It was found that there were logarithmic terms in the four point correlation function of the fundamental spin $\half$ primaries. This model was re-examined recently \cite{Nichols:2001du,Kogan:2001nj} and we shall follow up on a few points which were studied there:  the algebra of the integer dimension fields and also the hamiltonian reduction procedure.

One of the interesting features of LCFT is the existence of a hidden symmetry which means there are extra fields with integer conformal dimensions \cite{Caux:1996nm}. In the case of the integrable representations these were previously studied in \cite{MST,ST} in which it was found that the rational solutions to the KZ equation were in one-one correspondence with the extensions of the chiral algebra. In the $SU(2)_0$ model the integrable sector is trivial (the affine current $J^a$ is a null vector) and one is forced to immediately consider non-integrable representations.

In this article we wish to demonstrate that the $SU(2)_0$ WZNW model possesses several chiral algebras, beyond the normal affine Lie algebra, that are generated by the integer dimension fields. We shall construct the lowest lying currents directly in the the free  field approach and shall discover that the fields naturally become extended multiplets in exact analogue to the $c=-2$ model. We shall then demonstrate how such an extra extended multiplet structure emerges from the rational solutions of the KZ equation (The $j=1$ solutions are a special case of a more general solution recently found in \cite{Hadjiivanov:2001kr}). We shall finally show that upon Hamiltonian reduction these multiplets become precisely those previously found in the $c=-2$ triplet model. This strongly suggests that such currents could be used to construct a quasi-rational $SU(2)_0$ theory in exact analogue to the triplet model of $c=-2$ \cite{Gaberdiel:1998ps}. We shall briefly comment on the expected structure of the general extended $SU(2)_k$ models for $k\in N$.

\section{Extended Chiral Algebras}
\subsection{Free field representation}
We shall use the standard Wakimoto representation \cite{Wakimoto:1986gf} for the affine Lie algebra $\widehat{SU(2)}$ specialised to the case $k=0$. Our currents are:
\bea
J^+=\beta ~~~~~~
J^3=i\p \phi +\gamma \beta ~~~~~~
J^-=-2i \p \phi \gamma - \beta\gamma^2
\eea
where $\beta$ and $\gamma$ obey the standard free field relations:
\be
\beta(z) \beta(w) \sim 0 \sim \gamma(z) \gamma(w) ~~~~ \beta(z) \gamma(w) \sim \f{-1}{z-w} \sim -\gamma(z) \beta(w)
\ee
and $\phi$ is a free boson:
\bea
\phi(z) \phi(w) \sim - \ln(z-w)
\eea
these obey the standard affine Lie algebra with zero central extension:
\be \label{eqn:SU2KM}
J^3(z) J^{\pm}(w) \sim \pm \f{J^{\pm}(w)}{z-w} ~~~ J^+(z)J^-(w) \sim \f{2 J^3(w)}{z-w} ~~~ J^3(z)J^3(w) \sim 0
\ee
The stress tensor is just the standard Sugawara one:
\bea \label{eqn:sugawara}
T&=&\f{1}{2}\left( \half J^+J^- +\half J^-J^+ + J^3J^3 \right) 
= -\beta \p \gamma -\f{1}{2}\p \phi \p \phi + \f{i}{2}\p^2 \phi
\eea
We now observe that the $\phi$ part of this is none other than the $c=-2$ system. This fact will be essential in order to see the extended multiplet nature in the free field representations. We can thus write:
\bea
\xi \sim e^{i \phi} ~~~~~~~ \eta \sim e^{-i \phi} 
\eea
where $\xi,\eta$ are the symplectic fermion system:
\bea
\xi(z)\eta(w) \sim \f{1}{z-w} \sim \eta(z)\xi(w) ~~~~~~~ i \p \phi =  \xi \eta 
\eea
The currents and stress tensor become:
\bea \label{eqn:cminus2}
J^+&=&\beta ~~~~~~
J^3=\xi \eta + \beta \gamma  ~~~~~~
J^-=-2 \xi \eta \gamma - \beta\gamma^2 \nonumber\\
\\
T&=& -\beta \p \gamma - \xi \p \eta \nonumber
\eea
The affine Lie algebra primary fields $\Phi$ obey:
\bea \label{eqn:KMprimary}
J^a(z) \Phi(w) \sim \f{t^a \Phi(w)}{z-w}
\eea
and have conformal dimension:
\bea
h=\f{j(j+1)}{2}
\eea
Normally the vertex operators of primary affine Lie algebra fields are given by:
\bea
\Phi_{j,m}=(-\gamma)^{j-m}e^{ij\phi}
\eea
 However in order to calculate correlators with such operators vacuum charges and screening operators must be inserted \cite{Dotsenko:1990ui,Dotsenko:1991zb}. This is in general quite complicated moreover the result is in general only an integral representation. As we shall see for certain operators in $SU(2)_0$ one can obtain correlators and OPEs in a much simpler way.
\subsection{Extended algebras}
For $j \in N$ we have $h \in N$ and so one may suspect that there is a simpler way of evaluating correlators for such fields. We found the following expressions for the fields \footnote{We are extremely grateful to K. Thielemans for giving us a copy of his OPEDEFS Mathematica program \cite{Thielemans:1991uw} which was invaluable in finding these expressions.}. For brevity we shall only write the highest component, that is the one annihilated by $J^+$ . All others in the multiplet may be obtain by the action of $J^-$ (for $j=2$ we list the full multiplets in the Appendix):
\begin{itemize}
\item{$j=1$}
\bea \label{eqn:jequalone}
J^+=\beta
\eea
\item{$j=2$}
\bea \label{eqn:jequaltwo}
&&G^{++}=-\beta \p \xi + \p \beta \xi\\
&&H^{++}=\beta \beta \beta \eta
\eea
\item{$j=3$}
\bea
X^{+++}&=&\beta \beta \beta \beta \beta \p \eta \eta \\
W^{+++}&=&\f{1}{2} \beta \beta \beta \eta \p^2 \xi - \f{1}{2} \beta \beta \beta \p \eta \p \xi - \f{3}{2} \p \beta \beta \beta \eta \p \xi  +\f{1}{2} \p \beta \beta \beta \p \eta \xi + \f{3}{2} \p \beta \p \beta \beta \eta \xi \nonumber \\
&& \quad + \p \beta \p \beta \p \beta - \f{1}{2} \p^2 \beta \beta \beta \eta \xi -\f{3}{4} \p^2 \beta \p \beta \beta + \f{1}{12} \p^3 \beta \beta \beta \\
Y^{+++}&=&-\beta \p^2 \xi \p \xi + \p \beta \p^2 \xi \xi - \p^2 \beta \p \xi \xi
\eea
\end{itemize}
It is a non-trivial fact that these actually belong to the \emph{discrete} representations of $SU(2)$ and thus the lowest component in the multiplet is annihilated by $J^-$.

We have also included the $j=1$ field (\ref{eqn:jequalone}) which is of course just the affine current $J^+$ itself. We also found other solutions $\xi,\p \xi \xi, \p^2 \xi \p \xi \xi$ for $j=1,2,3$ respectively. They are in fact extra primary operators that have different behaviour under crossing symmetries. For instance the extra $j=1$ field is one of the fermionic operators found in \cite{Nichols:2001du}. We shall not discuss these operators here as these are not responsible for the rational solutions to the KZ equation that we shall discuss later.

To summarize: The use of the symplectic fermion system (\ref{eqn:cminus2}) produces exactly the correct chiral field content of the $SU(2)_0$ model (We have checked this explicitly up $h=6$ operators). We shall find that these chiral fields are also in one-one correspondence with the rational solutions of the KZ equation.
\subsection{Operator Product Expansions}
We can immediately observe that $j=1,2,3$ are alternately bosonic, fermionic, bosonic and that there are respectively a singlet, doublet and a triplet of possible fields. We remind the reader once again that each of the two solutions for $j=2$ yields a full $SU(2)$ multiplet with $m=+2,\cdots m=-2$ and that we have only written the $m=+2$ term. Although the full OPEs are quite complicated (see Appendix for full $j=2$ algbera) we see that by using the above symmetries and the fact that a chiral algebra is completely determined by its the singular terms we can already deduce almost all the structure. We know:
\bea
[j=0] \times [j=J]=[j=J]
\eea
where $[~]$ denotes operators and all their affine Lie algebra descendents (In particular the $j=1$ state is of course just a descendent of the identity). A special case of this is (\ref{eqn:KMprimary}).

Now we shall find it convenient to use indices $\alpha,\beta$ for extra doublet index structure and $a,b,c$ for the triplet. We now want to analyse the OPE of $[j=2]^{\alpha}$ with itself. Clearly as these are both fermionic the result must be bosonic. Furthermore the $j=3$ operators have $h=6$ and so cannot occur in the singular terms. Thus we must have:
\bea
[j=2]^{\alpha} \times [j=2]^{\beta}=d^{\alpha \beta} [j=0]
\eea
We have explicitly calculated the entire algebra of the $[j=2]^{\pm}$ operators (see Appendix) and have verified that it is indeed of the above form. 

By similar arguments we find:
\bea
[j=2]^{\alpha} \times [j=3]^{a}=t^{a \alpha}_{\beta} [j=2]^{\beta}
\eea
In this case we can only produce fermionic operators in the singular terms with dimension at most  $3+6=9$. In particular this excludes the fermionic $j=4$ operator with $h=10$. Similarly:
\bea
[j=3]^{a} \times [j=3]^{b}=g^{ab} [j=0] + f^{ab}_c [j=3]^c
\eea
We thus expect $j=1,2,3$ to yield a closed W-algebra. We also expect that in the absence of other symmetries, such as orbifolding, we cannot get a finite W-algebra involving $j > 3$.

\section{Knizhnik-Zamolodchikov equation}

\subsection{Auxiliary variables}

\label{sec:2and3pt}

It will be convenient in much of this paper to introduce the following representation for the  $SU(2)$ generators \cite{Zamolodchikov:1986bd}:
\bea \label{eqn:repn}
J^+=x^2\frac{\p}{\p x}-2jx, ~~~ 
J^-=-\frac{\p}{\p x}, ~~~
J^3=x\frac{\p}{\p x}-j \eea
There is also a similar algebra in terms of $\bar{x}$ for the antiholomorphic part. It is easily verified that these obey the global $SU(2)$ algebra.

We introduce primary fields, $\phi_j(x,z)$  of the affine Lie algebra:
\be
J^a(z)\phi_j(x,w) = \frac{1}{z-w} J^a(x) \phi_j(x,w) 
\ee
where $J^a(x)$ is given by (\ref{eqn:repn}). The fields $\phi_j(x,z)$ are also primary with respect to the Virasoro algebra with $L_0$ eigenvalue:
\be
h=\frac{j(j+1)}{k+2}
\ee
The two point functions are fully determined using global $SU(2)$ and conformal transformations and can be normalised in the standard way:
\be \label{eqn:2pt}
\la \phi_{j_1}(x_1,z_1) \phi_{j_2}(x_2,z_2) \ra = \delta^{j_1}_{j_2}x_{12}^{2j_1}z_{12}^{-2h}
\ee
The general form of the three point function is:
\bea \label{eqn:3pt}
\la \phi_{j_1}(x_1,z_1) \phi_{j_2}(x_2,z_2) \phi_{j_3}(x_3,z_3) \ra = C(j_1,j_2,j_3)~~ x_{12}^{j_1+j_2-j_3} x_{13}^{j_1+j_3-j_2} x_{23}^{j_2+j_3-j_1} \\
z_{12}^{-h_1-h_2+h_3} z_{13}^{-h_1-h_3+h_2} z_{23}^{-h_2-h_3+h_1} \nonumber
\eea
The $C(j_1,j_2,j_3)$ are the structure constants which in principle completely determine the entire theory.

For the case of the four point correlation functions of $SU(2)$ primaries the form is determined by global conformal and $SU(2)$ transformations up to a function of the cross ratios.
\bea \label{eqn:correl} 
\langle \phi_{j_1}(x_1,z_1) \phi_{j_2}(x_2,z_2) \phi_{j_3}(x_3,z_3) \phi_{j_4}(x_4,z_4) \rangle
&=&z_{43}^{h_2+h_1-h_4-h_3}z_{42}^{-2h_2}z_{41}^{h_3+h_2-h_4-h_1} \nonumber \\
& & z_{31}^{h_4-h_1-h_2-h_3}x_{43}^{-j_2-j_1+j_4+j_3}x_{42}^{2j_2}  \\
& & x_{41}^{-j_3-j_2+j_4+j_1}x_{31}^{-j_4+j_1+j_2+j_3}~F(x,z) \nonumber
\eea 
Here the invariant cross ratios are:
\be 
x=\frac{x_{21}x_{43}}{x_{31}x_{42}} ~~~ z=\frac{z_{21}z_{43}}{z_{31}z_{42}} 
\ee
\subsection{The Knizhnik-Zamolodchikov Equation}
Correlation functions of the WZNW model satisfy a set of partial differential equations known as  Knizhnik-Zamolodchikov (KZ) equation due to constraints from the null states following from (\ref{eqn:sugawara}). These are:
\be
|\chi \ra = ( L_{-1} - \frac{1}{k+2} \eta_{ab}J^a_{-1}J^b_0 ) |\phi \ra
\ee
For two and three point functions this gives no new information. However for the four point function (\ref{eqn:correl}) it becomes a partial differential equation for $F(x,z)$. For  a compact Lie group this equation can be solved \cite {Knizhnik:1984nr} as it reduces to a set of ordinary differential equations.
\be
\left[(k+2) \frac{\p}{\p z_i}+\sum_{j\neq i}\frac{\eta_{ab} J^a_i \otimes J^b_j}{z_i-z_j} \right] \left<\phi_{j_1}(z_1) \cdots \phi_{j_n}(z_n) \right> =0 
\ee
where $k$ is the level of the $SU(2)$ WZNW model. 

If we now use our representation (\ref{eqn:repn}) we find the KZ equation for four point functions is:
\be \label{eqn:KZ}
(k+2) \frac{\p}{\p z} F(x,z)=\left[ \frac{\CP}{z}+\frac{\CQ}{z-1} \right] F(x,z)
\ee
Explicitly these are:
\bea
\CP \!\!\!\!&=&\!\!-x^2(1-x)\frac{\p^2}{\p x^2}+((-j_1-j_2-j_3+j_4+1)x^2+2j_1x+2j_2x(1-x))\frac{\p}{\p x} \nonumber \\
& & +2j_2(j_1+j_2+j_3-j_4)x-2j_1j_2 \\
\CQ \!\!\!\!&=&\!\!-(1-x)^2x\frac{\p^2}{\p x^2}-((-j_1-j_2-j_3+j_4+1)(1-x)^2+2j_3(1-x)+2j_2x(1-x))\frac{\p}{\p x} \nonumber \\
& & +2j_2(j_1+j_2+j_3-j_4)(1-x)-2j_2j_3 
\eea
\section{Rational Solutions to the $SU(2)_0$ KZ equation}
For the case of $j \in N$ we found for all cases studied that we have either rational or logarithmic solutions. As we are discussing chiral algebras we shall concentrate on the rational solutions.

We can write the general solution to the KZ equation as: 
\be
F(x,z)=F_0(z)+xF_1(z)+x^2F_2(z)+ \cdots +x^{2j}F_{2j}(z)
\ee
In \cite{Kogan:2001nj} an ansatz was found for the four point correlator of the spin $j$ operator:
\bea
F_0(z)=z^{-j(j-1)}(1-z)^{-j(j+1)}F(j,-j;1;z)
\eea
For $j \in N$ these are single-valued and using the crossing symmetries one can easily find a set of rational solutions on which one can perform the conformal bootstrap.

For $j=1$ we find only one solution:
\bea \label{eqnjone}
F(x,z)=-\f{1}{2(z-1)}+\f{x}{z}+\f{x^2}{2z(z-1)}
\eea
This obeys:
\bea
F(1-x,1-z)&=&F(x,z) \\
z^{-2h}x^{2j}F\left(\f{1}{x},\f{1}{z}\right)&=&F(x,z)
\eea
If we wish to identify this with the affine current $J^a$ then we have a problem as from the affine Lie algebra (\ref{eqn:SU2KM}) we get:
\bea
\left< J^a(z_1) J^b(z_2) J^c(z_3) J^d(z_4) \right>=0
\eea
The resolution is that there should be some field $N^a$ (this was found in \cite{Caux:1997kq}) which is a logarithmic partner to $J^a$ and thus:
\bea
\left< J(x_1,z_1) J(x_2,z_2) J(x_3,z_3) N(x_4,z_4) \right>=\f{x_{42}^{2}x_{31}^{2}F(x,z)}{z_{42}^{2}z_{31}^{2}}
\eea
We still have full crossing symmetry and expanding this will reveal the correct $JJ$ OPEs. We need not explicitly discuss $N^a$ other than assume that there exists such a field which makes the four point functions non-zero so that we can use the correlators to obtain the OPEs.

For $j=2$ we find two solutions:
\bea \label{eqn:jtwo}
F_1(x,z)&=&\f{1}{z^5}\left( (-z^3-2z^4)+(-12z^2-16z^3+4z^4)x+(-18z+18z^3)x^2 \right. \nonumber  \\
&&\left.+(-4+16z+12z^2)x^3+(2+z)x^4 \right) \\
F_2(x,z)&=&\f{1}{(z-1)^5} \left( (10z-15z^2+9z^3-2z^4)+(60-140z+108z^2-36z^3+4z^4)x \right. \nonumber\\
&&\left.+(-90+162z-90z^2+18z^3)x^2+(36-44z+12z^2)x^3+(-3+z)x^4 \right) \nonumber
\eea
These obey:
\bea
F_1(1-x,1-z)&=&F_2(x,z)\nonumber \\
F_2(1-x,1-z)&=&F_1(x,z)\\
x^4z^{-6}F_1\left(\f{1}{x},\f{1}{z} \right)&=&-F_1(x,z) \nonumber\\
x^4z^{-6}F_2\left(\f{1}{x},\f{1}{z} \right)&=&-F_1(x,z)+F_2(x,z) \nonumber
\eea
We can already see the fermionic nature of these solutions as we expected.

For $j=3$ we have three solutions (see Appendix). They exhibit a triplet nature  as we expected. All these correlators agree with the leading forms of the OPEs from the free field representations. The subleading terms are technically much more complicated to verify fully.

The correlator $ \left< 1112 \right>$ has two rational solutions $1-z,\f{1}{1-z}$. However neither of these leads to a chiral correlator respecting the full crossing symmetries if the $j=1$ operators are all identical. This was to be expected as the fermionic nature of the chiral $j=2$ operator means this correlator must vanish.
\section{Hamiltonian reduction}
There is another interesting way in which $SU(2)_0$ is related to the $c=-2$ theory. When we do a quantum hamiltonian reduction of $SU(2)_k$ theories normally by imposing the constraint $J^+ \sim 1$ it is well known \cite{Drinfeld:1984qv,Polyakov:1988qz,Alekseev:1989ce,Bershadsky:1989mf,Feigin:1990pn} that we get the $c_{k+2,1}$ minimal models. 

The central charge and conformal weights of the reduced theory are given by:
\bea
c=c_{k+2,1}=13-6 \left( k+2+\f{1}{k+2} \right) 
\eea
\bea  \label{eqn:reduced}
h=\f{j(j+1)}{k+2}-j \
\eea
Thus $k=0$ reduces to $c_{2,1}=-2$.
Here we follow an elegant realisation of this reduction by Petkova et al. \cite{Furlan:1993mm,Ganchev:1992af,Ganchev:1993ci} that allow us to perform this at the level of the correlation functions. The motivation for such a procedure comes from the observation that setting $x=z$ in two and three point functions (\ref{eqn:2pt},\ref{eqn:3pt}) gives the expected results in the reduced theory with the correct conformal weight (\ref{eqn:reduced}).

Several examples of this were demonstrated in \cite{Kogan:2001nj} for the half-integer spin cases involving logarithms. Here we shall instead discuss the chiral solutions of the integer spin operators. For the case of $k=0$ we get:
\bea
\ba{lllllll}
j~~~~ 0 &  1 &  2  & 3 \nonumber \\
h~~~~ 0 & 0 & 1  & 3 \nonumber
\ea
\eea
Now performing such a reduction on the $j=1$ correlator (\ref{eqnjone}) we find:
\bea
F(x,z) \rightarrow 1
\eea
Such a simple correlator is due to the fact that the $j=1$ field reduces to an $h=0$ field. This does not generate a chiral algebra in the reduced theory as it is just the identity.

For the $j=2$ correlators (\ref{eqn:jtwo}) we get:
\bea
F_1(x,z) &\rightarrow& 1- \f{1}{z^2} \\
F_2(x,z) &\rightarrow& 1- \f{1}{(1-z)^2} 
\eea
These reproduce precisely the conformal blocks of the $h_{1,2}=1$ operators in the $c=-2$ theory as claimed. The doublet nature is unaffected by the reduction. Thus the Hamiltonian reduction of the multiplets $G,H$ (\ref{eqn:jequaltwo}) gives precisely the fields $\chi^{\pm}$ which are the fermion doublet in $c=-2$:
\bea
\chi{\pm}(z) \chi{\pm}(w) \sim 0 ~~~~~ \chi^+(z) \chi^-(w) \sim \f{1}{(z-w)^2}
\eea
The symplectic fermion algebra is clearly automatically associative by which we mean that the Jacobi identity is automatically satisfied given the chiral algebra. However for the $j=2$ algebra in $SU(2)_0$ we found that it was only associative \textbf{provided} that certain null vectors decouple. For example:
\bea
{\mathcal N} =4 J^3 G^{++}+J^{+}G^{+}-2 \p G^{++}
\eea
It can be verified that this decouples, as it must do, using the free field representation of the operators given earlier. However from the algebraic point of view the algebra is only consistent if such null vectors decouple. We hope this may give some insight into how to form a quasi-rational $SU(2)_0$ theory.

Reduction of the $j=3$ correlators (\ref{eqn:jthree})  gives:
\bea
F_1 &\rightarrow& \f{1}{(z-1)^6} z^4 \left(6-6z+z^2 \right)\\
F_2 &\rightarrow& \f{1}{z^6(z-1)^6} \left( 2-12z+12z^2+50z^3-225z^4+468z^5-588z^6+468z^7-225z^8 \right.\nonumber\\
&&\left.+50z^9+12z^{10}-12z^{11}+2z^{12} \right)\\
F_3 &\rightarrow& \f{1}{z^6} \left( 1-9z^2+16z^3-9z^4+z^6 \right)
\eea
These again, as expected, reproduce the conformal blocks for $h_{1,3}=3$ operators in $c=-2$. More explicitly by analysing the pole structure and symmetries we find that these are related (up to normalisation) in the following way to the standard fields \cite{Kausch:1995py}:
\bea
F_1 &\rightarrow& \left< W^+(0) W^+(z) W^-(1) W^-(\infty) \right> \nonumber \\
F_2 &\rightarrow& \left< W^3(0) W^3(z) W^3(1) W^3(\infty) \right> \\
F_3 &\rightarrow& \left< W^+(0) W^-(z) W^-(1) W^+(\infty) \right>\nonumber
\eea
The first of these is the unique solution that has no singular terms as $z \rightarrow 0$. The third is similar but with no singular terms as $z \rightarrow 1$. The second is the unique solution invariant under all crossing symmetries.

 Analysing these is sufficient to fully reconstruct the algebra generated by the $W^a$ fields \cite{Kausch:1995py}:
{\small
\bea
T(z) T(w) &\sim& \f{-2}{2 (z-w)^4} + \f{2 T(w)}{(z-w)^2} + \f{\p T(w)}{z-w} \nonumber \\
T(z) W^a(w) &\sim& \f{3 W^a(w)}{(z-w)^2}+ \f{\p W^a(w)}{z-w} \\
W^a(z) W^b(w) &\sim&  g^{ab} \left( \f{1}{(z-w)^6}- 3 \f{T(w)}{(z-w)^4}-\f{3}{2} \f{\p T(w)}{(z-w)^3} + \f{3}{2} \f{\p^2 T(w)}{(z-w)^2} -4 \f{(T^2)(w)}{(z-w)^2} \right. \nonumber  \\
&&~~~~ \left. + \f{1}{6} \f{\p^3 T(w)}{z-w} -4 \f{\p(T^2)(w)}{z-w} \right) \nonumber \\ 
&& - 5 f^{ab}_c \left( \f{W^c(w)}{(z-w)^3} + \f{1}{2}\f{\p W^a(w)}{(z-w)^2} + \f{1}{25} \f{\p^2 W^c(w)}{z-w} + \f{1}{25} \f{(TW^c)(w)}{z-w} \right) \nonumber
\eea}
The diagonal generator $W^3$ generates an automatically associative W-algebra that is precisely the Zamolodichikov $W(2,3)$ algebra \cite{Zamolodchikov:1985wn} at $c=-2$. The full triplet algebra however is only associative if certain null vectors decouple \cite{Gaberdiel:1996np}.  We have shown that the above algebra \textbf{can} come from a hamiltonian reduction of an $SU(2)$ structure. This is in complete contrast to the normal case of Hamiltonian reduction of the $SU(2)$ affine Lie algebra which leads to the Virasoro algebra only.
\subsection{General $SU(2)_k$ structure}
We shall now comment briefly on the structure that we expect in $SU(2)_k$ for $k \in N$. We have in general an affine Lie algebra null vector that transforms in the $j=k+1$ representation. Imposing this as a null vector on states gives the normal rational affine Lie algebra theory with $0 \le j \le \f{k}{2}$. For $SU(2)_0$ this null vector was precisely the affine current $J^a$ and this rational set was trivial. Using (\ref{eqn:reduced}) we see that the reduction of this null vector gives an $h=0$ state in the $c_{k+2,1}$ theory. We thus see that extending the model by adding a logarithmic partner which prevents the affine Lie algebra null vector decoupling automatically leads to a logarithmic partner for the vacuum in the $c_{p,1}$ models.

All the $c_{p,1}$ models have a triplet algebra generated by the $h_{1,3}$ fields \cite{Kausch:1991vg}. These come from the hamiltonian reduction of $j=2k+3$ fields in $SU(2)_k$. We therefore conjecture that there will also be a triplet algebra generated by these fields and correspondingly, in general, three rational solutions to the KZ equation. However there remain many subtleties in the reduction process \cite{Petersen:1996xn}. In the case of $c=-2$ there was also the $h_{1,2}=1$ fields which generated a doublet algebra but this is not typical of the other $c_{p,1}$ models.
\section{Conclusion}
We have presented evidence, from both the free field construction and the explicit correlation functions, that there exist extended algebras in the $SU(2)_0$ model. Their extra multiplet structure is identical to that found in $c=-2$. We have shown that expressing the $c=-2$ part of the $SU(2)_0$ stress tensor in terms of symplectic fermions leads us to \emph{exactly} the correct chiral operator content of the model. There is a chiral doublet of fermionic $j=2$ fields and a triplet of bosonic $j=3$ fields. Moreover hamiltonian reduction of these fields gives us precisely the $h=1$ doublet and $h=3$ triplet in $c=-2$. It is still an open question as to what extent the reduction extends to other operators of the theory. 

In view of the fact that the triplet algebra in $c=-2$ yielded a quasi-rational model we may very much hope that these extended algebras will do the same for $SU(2)_0$. We have verified that the $j=2$ algebra is only consistent if certain null vectors decouple from the spectrum. It would be very interesting to be able to produce a full rational model of $SU(2)_0$  as this may be useful in shedding light on the many questions that remain on the general structure at $c=0$\cite{WorkinProgress}.

There are clearly many outstanding questions. It would be very interesting to investigate other $\widehat{SU(2)}$ theories and in particular the case $k \in N$ which we have already mentioned that we expect to have a similar structure to $c_{k+2,1}$. For $k \ne N$ the extended structure remains an interesting open problem.
\section{Acknowledgements}
I would like to thank I. I. Kogan for useful and stimulating discussions. I have received funding from the Martin Senior Scholarship awarded by Worcester College, Oxford. 
\appendix
\section{Appendix}
\subsection{Correlation functions}
The correlation functions for the $j=3$ operators are:
\bea \label{eqn:jthree}
W_1(x,z)&=&\f{1}{462(z-1)^{11}}z^3 \left\{ (-252z+756z^2-910z^3+560z^4-190z^5+36z^6-3z^7)\right. \nonumber \\ 
&&+(-2520+9576z-14532z^2+11340z^3-4860z^4+1140z^5-138z^6+6z^7)x\nonumber\\
&&+(6300-21000z+27510z^2-18000z^3+6150z^4-1050z^5+75z^6)x^2\\
&&+(-5600+16240z-17840z^2+9200z^3-2200z^4+200z^5)x^3\nonumber\\
&&+(2100-5010z+4200z^2-1425z^3+150z^4)x^4\nonumber \\
&&\left.+(-300+516z-258z^2+30z^3)x^5+(10-8z+z^2)x^6 \right\}\nonumber
\\
W_2(x,z)&=&\f{1}{231z^{11} (z-1)^{11}} \nonumber\\
&&\left\{ (-z^5+8z^6-10z^7-70z^8+455z^9-1456z^{10}+ 3003z^{11}-4290z^{12}+4290z^{13} \right.\nonumber\\
&&-2860z^{14}+875z^{15}+560z^{16}-910z^{17}+560z^{18}-190z^{19}+36z^{20}-3z^{21})\nonumber\\
 &&+ (-30z^4+258z^5-516z^6-1380z^7+11970z^8-40950z^9+88998z^{10}\nonumber\\
&&-134706z^{11}+145860z^{12}-111540z^{13}+54510z^{14}-7770z^{15}-12180z^{16}\nonumber\\
&&+11340z^{17}-4860z^{18}+1140z^{19}-138z^{20}+6z^{21})x\nonumber\\
&&+(-150z^3+1425z^4-4200z^5-870z^6+46200z^7-184275z^8+436800z^9\nonumber\\
&&-718575z^{10}+858000z^{11}-750750z^{12}+471900z^{13}-194775z^{14}+30450z^{15}\nonumber\\
&&+21630z^{16}-18000z^{17}+6150z^{18}-1050z^{19}+75z^{20})x^2\nonumber\\
&&+(-200z^2+2200z^3-9200z^4+13920z^5+26880z^6-197400z^7+564200z^8\nonumber\\
&&-1058200z^9+1430000z^{10}-1430000z^{11}+1058200z^{12}-564200z^{13}\nonumber\\
&&+197400z^{14}-26880z^{15}-13920z^{16}+9200z^{17}-2200z^{18}+200z^{19})x^3\\
&&+(-75z+1050z^2-6150z^3+18000z^4-21630z^5-30450z^6+194775z^7-471900z^8\nonumber\\
&&+750750z^9-858000z^{10}+718575z^{11}-436800z^{12}+184275z^{13}-46200z^{14}\nonumber\\
&&+870z^{15}+4200z^{16}-1425z^{17}+150z^{18})x^4\nonumber\\
&&+ (-6+138z-1140z^2+4860z^3-11340z^4+12180z^5+7770z^6-54510z^7\nonumber\\
&&+111540z^8-145860z^9+134706z^{10}-88998z^{11}+40950z^{12}-11970z^{13}\nonumber\\
&&+1380z^{14}+516z^{15}-258z^{16}+30z^{17})x^5\nonumber\\
&&+(3-36z+190z^2-560z^3+910z^4-560z^5-875z^6+2860z^7\nonumber\\
&&\left.-4290z^8+4290z^9-3003z^{10}+1456z^{11}-455z^{12}+70z^{13}+10z^{14}-8z^{15}+z^{16})x^6 \right\}\nonumber
\\
W_3(x,z)&=&\f{1}{462 z^{11}}(z-1)^3 \left\{ (-z^5-6z^6-3z^7)+(-30z^4-162z^5-54z^6+6z^7)x \right.\nonumber\\
&&+(-150z^3-675z^4+75z^6)x^2+(-200z^2-600z^3+600z^4+200z^5)x^3\\
&&\left.+(-75z+675z^3+150z^4)x^4+(-6+54z+162z^2+30z^3)x^5+(3+6z+z^2)x^6 \right\}\nonumber
\eea
\subsection{Operator Product Expansions}
Using the free field representation of the the affine Lie algebra generators and those of the $j=2$ representations $G$ and $H$ we can explicitly find the full algebra. The top component $G^{++}$ was given in the text (\ref{eqn:jequaltwo}). By operating repeatedly by $J^-$ (in an obvious notation) one gets:
\bea
G^{++}&=&- \beta \p \xi+ \p \beta \xi \nonumber\\
G^{+}&=&4 \beta \gamma \p \xi+4 \eta \p \xi \xi-4 \p \beta \gamma \xi \nonumber\\
G^0&=&-12  \beta \gamma \gamma \p \xi-24  \gamma \eta \p \xi \xi +12 \p \beta \gamma \gamma \xi \\
G^{-}&=& 24  \beta \gamma \gamma \gamma \p \xi+72 \gamma \gamma \eta \p \xi \xi-24 \p \beta \gamma \gamma \gamma \xi \nonumber\\
G^{--}&=&-24  \beta \gamma \gamma \gamma \gamma \p \xi -96  \gamma \gamma \gamma \eta \p \xi \xi + 24 \p \beta \gamma \gamma \gamma \gamma \xi \nonumber
\eea
The $H$ multiplet is given by:
\bea
H^{++}&=&\beta \beta \beta \eta \nonumber \\
H^{+}&=& -4  \beta \beta \beta \gamma \eta+6 \beta \beta \p \eta-6 \p \beta \beta \eta \\
H^{0}&=&12 \beta \beta \beta \gamma \gamma \eta-36 \beta \beta \gamma \p \eta+36 \beta \p \eta \eta \xi+6 \beta \p^2 \eta+36 \p \beta \beta \gamma \eta-24 \p \beta \p \eta+6 \p^2 \beta \eta \nonumber\\
H^{-}&=&-24  \beta \beta \beta \gamma \gamma \gamma \eta+108 \beta \beta \gamma \gamma \p \eta-216 \beta \gamma \p \eta \eta \xi-36 \beta \gamma \p^2 \eta-108 \p \beta \beta \gamma \gamma \eta \nonumber\\
&& \quad +144 \p \beta \gamma \p \eta-144 \p \eta \eta \p \xi-36 \p^2\beta \gamma \eta+36 \p^2 \eta \eta \xi \nonumber\\
H^{--}&=&24  \beta \beta \beta \gamma \gamma \gamma \gamma \eta-144  \beta \beta \gamma \gamma \gamma \p \eta+432 \beta \gamma \gamma \p \eta \eta \xi+72 \beta \gamma \gamma \p^2 \eta+576 \gamma \p \eta \eta \p \xi \nonumber\\
&& \quad -144 \gamma \p^2 \eta \eta \xi+144 \p \beta \beta \gamma \gamma \gamma \eta-288 \p \beta \gamma \gamma \p \eta+72 \p^2 \beta \gamma \gamma \eta \nonumber
\eea
One can verify that:
\bea
G^{a}(z) G^{b}(w) &\sim& 0 \\
H^{a}(z) H^{b}(w) &\sim& 0 
\eea
so the only non-trivial OPEs are between $G$ and $H$ multiplets. 
%
\bea
G^{++}(z) H^{++}(w) &\sim& \f{ J^+ J^+ J^+ J^+ }{(z-w)^2}+ \f{2  \p J^+ J^+ J^+ J^+}{(z-w)}
\\
G^{++}(z) H^{+}(w) &\sim&\f{12  J^+ J^+ J^+}{(z-w)^3}+ \f{-4 J^+ J^+ J^+ J^3+12 \p J^+ J^+ J^+ }{(z-w)^2} \\
&&+ \f{1}{z-w} \left( -4 J^+ J^+ J^+ J^3 J^3-4 J^+ J^+ J^+ J^+ J^- -4 J^+ J^+ J^+ \p J^3 \right. \nonumber\\
&& \quad \left. -12 \p J^+ \p J^+ J^++12 \p^2 J^+ J^+ J^+ \right)\nonumber
\\
G^{++}(z) H^{0}(w) &\sim& \f{72 J^+ J^+}{(z-w)^4}+ \f{-72 J^+ J^+ J^3}{(z-w)^3} \nonumber\\
&&+\f{-12 J^+ J^+ J^+ J^--72 \p J^+ J^+ J^3-72 \p J^+ \p J^++36 \p^2 J^+ J^+ }{(z-w)^2} \\
&&+ \f{1}{z-w} \left( 24 J^+ J^+ J^3 J^3 J^3+24 J^+ J^+ J^+ J^3 J^- -12 J^+ J^+ J^+ \p J^- \right. \nonumber\\
&& \quad \left.  -12 J^+ J^+ \p^2 J^3  +36 \p J^+ J^+ J^+ J^-+72 \p J^+ J^+ \p J^3 \right. \nonumber\\
&& \quad \left. -72 \p^2 J^+ J^+ J^3 -36 \p^2 J^+ \p J^+ +12 \p^3 J^+ J^+ \right)\nonumber\\
G^{++}(z) H^{-}(w) &\sim& \f{144 J^+}{(z-w)^5}+ \f{-432 J^+ J^3-144 \p J^+}{(z-w)^4}\nonumber\\
&&+ \f{144 J^+ J^3 J^3-72 J^+ J^+ J^-+144 J^+ \p J^3-288 \p J^+ J^3}{(z-w)^3} \nonumber \\
&&+ \f{1}{(z-w)^2} \left( 48 J^+ J^3 J^3 J^3+72 J^+ J^+ J^3 J^-+24 J^+ \p^2 J^3 \right.\nonumber\\
&& \left. +144 \p J^+ J^3 J^3+288 \p J^+ \p J^3-216 \p^2 J^+ J^3\right) \\
&&+ \f{1}{z-w} \left(-54 J^+ J^3 J^3 J^3 J^3-36 J^+ J^+ J^3 J^3 J^-+108 J^+ J^+ J^3 \p J^- \right. \nonumber\\
&&\quad +18 J^+ J^+ J^+ J^- J^- -72 J^+ J^+ \p J^3 J^-+36 J^+ \p J^3 J^3 J^3 \nonumber\\
&& \quad -90 J^+ J^+ \p J^3 \p J^3+72 J^+ \p^2 J^3 J^3-6 J^+ \p^3 J^3 +24  \p J^+ J^3 J^3 J^3  \nonumber\\
&&\quad -108  \p J^+ J^+ J^3 J^- +108 \p J^+ J^+ \p J^- -72 \p J^+ \p J^3 J^3 +24 \p J^+ \p^2 J^3 \nonumber\\
&& \quad \left. +108 \p^2 J^+ J^3 J^3-108 \p^2 J^+ J^+ J^- +108 \p^2 J^+ \p J^3-72 \p^3 J^+ J^3 \right) \nonumber\\
G^{++}(z) H^{--}(w) &\sim& \f{-576 J^3}{(z-w)^5}+ \f{288 J^3 J^3-576  J^+ J^-+288 \p J^3}{(z-w)^4}  \nonumber\\
&&+ \f{288 J^+ J^3 J^-+288 J^+ \p J^- -576 \p J^+ J^-}{(z-w)^3}\nonumber \\
&&+ \f{1}{(z-w)^2} \left( -48 J^3 J^3 J^3 J^3+144 J^+ J^3 \p J^- +72 J^+ J^+ J^- J^- \right.  \nonumber\\
&& \quad -144 J^+ \p J^3 J^--144 \p J^3 \p J^3+144 \p J^+ J^3 J^-+432 \p J^+ \p J^-   \nonumber\\
&& \quad \left. +96 \p^2 J^3 J^3-432 \p^2J^+ J^- \right) \\
&&+ \f{1}{z-w} \left(24  J^3 J^3 J^3 J^3 J^3-48 J^+ J^3 J^3 J^3 J^--288 J^+ J^3 J^3 \p J^- \right. \nonumber \\
&&\quad -72 J^+ J^+ J^3 J^- J^-+288 J^+ \p J^3 J^3 J^--144 J^+ \p J^3 J^-\nonumber \\
&&\quad +96 J^+ \p^2 J^3 J^--48 \p J^3 J^3 J^3 J^3+72 \p J^3 \p J^3 J^3+144  \p J^+ J^3 J^3 J^-\nonumber \\
&&\quad -144 \p J^+ \p J^3 J^--48 \p^2 J^3 J^3 J^3-48 \p^2 J^3 \p J^3+216 \p^2J^+ J^3 J^-\nonumber \\
&&\quad \left. +216 \p^2 J^+ \p J^- +24 \p^3 J^3 J^3-144 \p^3 J^+ J^- \right) \nonumber \\
G^{+}(z) H^{++}(w) & \sim &  \f{-12  J^+ J^+ J^+}{(z-w)^3}+ \f{-4 J^+ J^+ J^+ J^3 -24 \p J^+ J^+ J^+ }{(z-w)^2} \\
&& + \f{4 J^+ J^+ J^+ J^3 J^3+4 J^+ J^+ J^+ J^+ J^- -12 \p J^+ J^+ J^+ J^3 -18 \p^2 J^+ J^+ J^+}{(z-w)}\nonumber  \\
G^{+}(z) H^{+}(w) &\sim & \f{-72 J^+ J^+}{(z-w)^4}+ \f{-72 \p J^+ J^+}{(z-w)^3} \nonumber\\
&&+ \f{1}{(z-w)^2} \left(24 J^+ J^+ J^3 J^3+8 J^+ J^+ J^+ J^--24 J^+ J^+ \p J^3\right.\nonumber \\
&& \quad \left.+48 \p J^+ J^+ J^3 +48 \p J^+ \p J^+ -60 \p^2 J^+ J^+ \right) \\
&&+ \f{1}{z-w} \left( 4 J^+ J^+ J^+ \p J^- +24 J^+ J^+ \p J^3 J^3-12 J^+ J^+ \p^2 J^3 \right. \nonumber\\
&& \quad +24 \p J^+ J^+ J^3 J^3+12 \p J^+ J^+ J^+ J^- +24 \p J^+ \p J^+ J^3\nonumber \\
&& \quad \left. +24 \p^2 J^+ J^+ J^3 +36 \p^2 J^+ \p J^+ -24 \p^3 J^+ J^+ \right) \nonumber\\
G^{+}(z) H^{0}(w) & \sim & \f{-144  J^+}{(z-w)^5}+ \f{144 J^+ J^3}{(z-w)^4}+ \f{144 J^+ J^3 J^3-144 J^+ \p J^3 +432 \p J^+ J^3}{(z-w)^3} \nonumber\\
&&+ \f{-48 J^+ J^3 J^3 J^3+144 J^+ \p J^3 J^3-48 J^+ \p^2 J^3 +216 \p^2 J^+ J^3 }{(z-w)^2}\nonumber\\
&&+ \f{1}{z-w} \left( -42 J^+ J^3 J^3 J^3 J^3-36 J^+ J^+ J^3 J^3 J^-+36 J^+ J^+ J^3 \p J^- \right.\\
&& \quad +6 J^+ J^+ J^+ J^- J^--36 J^+ \p J^3 J^3 J^3-54 J^+ \p J^3 \p J^3 +120 J^+ \p^2 J^3 J^3 \nonumber\\
&&\quad -18 J^+ \p^3 J^3 -72 \p J^+ J^3 J^3 J^3-180 \p J^+ J^+ J^3 J^-+36 \p J^+ J^+ \p J^- \nonumber\\
&&\quad -72 \p J^+ \p J^3 J^3 -72 \p J^+ \p J^+ J^- +72 \p J^+ \p^2 J^3 +36 \p^2 J^+ J^3 J^3\nonumber\\
&&\quad \left. -36 \p^2 J^+ J^+ J^- -108 \p^2 J^+ \p J^3 +96 \p^3 J^+ J^3 \right)\nonumber\\
G^{+}(z) H^{-}(w) & \sim & \f{288 J^3}{(z-w)^5}+ \f{576 J^3 J^3+144 J^+ J^-}{(z-w)^4} \nonumber\\
&& + \f{-288 J^3 J^3 J^3+288 J^+ J^3 J^-+288 \p J^3 J^3+432 \p J^+ J^-}{(z-w)^3}\nonumber\\
&&+ \f{1}{(z-w)^2} \left( -48 J^3 J^3 J^3 J^3-144 J^+ J^3 J^3 J^-+144 J^+ \p J^3 J^- \right. \nonumber\\ 
&&\quad \left. -288 \p J^3 J^3 J^3 -432 \p J^3 \p J^3 +384 \p^2 J^3 J^3+216 \p^2 J^+ J^- \right) \nonumber\\
&&+ \f{1}{z-w} \left( 84 J^3 J^3 J^3 J^3 J^3 -24 J^+ J^3 J^3 J^3 J^- -432 J^+ J^3 J^3 \p J^- \right. \\
&&\quad -108 J^+ J^+ J^3 J^- J^- +288 J^+ \p J^3 J^3 J^- -216 J^+ \p J^3 \p J^- \nonumber\\
&&\quad +192 J^+ \p^2 J^3 J^- -72 \p J^3 J^3 J^3 J^3 +252 \p J^3 \p J^3 J^3 \nonumber\\
&&\quad +216 \p J^+ J^3 J^3 J^- -432 \p J^+ J^3 \p J^- -216 \p J^+ J^+ J^- J^- \nonumber\\
&&\quad +216 \p J^+ \p J^3 J^- -312 \p^2 J^3 J^3 J^3  -216 \p^2 J^3 \p J^3 \nonumber\\
&& \quad \left. +324 \p^2 J^+ J^3 J^- -108 \p^2 J^+ \p J^- +132 \p^3 J^3 J^3+144 \p^3 J^+ J^- \right) \nonumber\\
G^{+}(z) H^{--}(w) & \sim & \f{-576 J^-}{(z-w)^5}+ \f{1728 J^3 J^- +576 \p J^-}{(z-w)^4} \nonumber\\
&&+ \f{-576 J^3 J^3 J^--576 J^3 \p J^- +288 J^+ J^- J^- +1152 \p J^3 J^-}{(z-w)^3}\nonumber\\
&&+ \f{1}{(z-w)^2} \left(-192 J^3 J^3 J^3 J^- -576 J^3 J^3 \p J^- -288 J^+ J^3 J^- J^- \right. \nonumber\\
&& \quad \left. -1152 \p J^3 \p J^- +960 \p^2 J^3 J^- \right) \nonumber\\
&&+ \f{1}{z-w} \left(216 J^3 J^3 J^3 J^3 J^-+768 J^3 J^3 J^3 \p J^-\right. \\
&&\quad +144 J^+ J^3 J^3 J^- J^- -432 J^+ J^3 \p J^- J^- -72 J^+ J^+ J^- J^- J^- \nonumber\\
&&\quad +288 J^+ \p J^3 J^- J^-  -432 J^+ \p J^- \p J^- +216 J^+ \p^2 J^- J^- \nonumber\\
&&\quad -1008 \p J^3 J^3 J^3 J^- +288 \p J^3 J^3 \p J^-  +360 \p J^3 \p J^3 J^- \nonumber\\
&&\quad +432 \p J^+ J^3 J^- J^- -720 \p^2 J^3 J^3 J^- -528 \p^2 J^3 \p J^- \nonumber\\
&&\quad \left.+216 \p^2 J^+ J^- J^- +312 \p^3 J^3 J^- \right)\nonumber\\
G^{0}(z) H^{++}(w) & \sim &  \f{72 J^+ J^+}{(z-w)^4}+ \f{72 J^+ J^+ J^3+144 \p J^+ J^+}{(z-w)^3} \nonumber\\
&&+ \f{-12 J^+ J^+ J^+ J^- +72 J^+ J^+ \p J^3 +72 \p J^+ J^+ J^3+108 \p^2 J^+ J^+ }{(z-w)^2}\\
&&+ \f{1}{z-w} \left( -24 J^+ J^+ J^3 J^3 J^3-24 J^+ J^+ J^+ J^3 J^- +48 J^+ J^+ \p^2 J^3 \right.\nonumber\\
&&\quad \left.-72 \p J^+ J^+ J^+ J^-+72 \p^2 J^+ J^+ J^3 +48 \p^3 J^+ J^+ \right) \nonumber\\
G^{0}(z) H^{+}(w) & \sim & \f{144 J^+}{(z-w)^5}+ \f{144 J^+ J^3+144 \p J^+}{(z-w)^4} \nonumber\\
&&+ \f{-144 J^+ J^3 J^3+288 J^+ \p J^3 -288 \p J^+ J^3+72 \p^2 J^+}{(z-w)^3}\nonumber\\
&&+ \f{1}{(z-w)^2} \left(-48 J^+ J^3 J^3 J^3-144 J^+ \p J^3 J^3+168 J^+ \p^2 J^3 \right. \nonumber\\
&&\quad \left. -144 \p J^+ J^3 J^3-144 \p J^+ \p J^3 -144 \p^2 J^+ J^3+24 \p^3 J^+ \right)\nonumber\\
&&+ \f{1}{z-w} \left(42 J^+ J^3 J^3 J^3 J^3+36 J^+ J^+ J^3 J^3 J^- -36 J^+ J^+ J^3 \p J^- \right.\\
&&\quad -6 J^+ J^+ J^+ J^- J^- -108 J^+ \p J^3 J^3 J^3+54 J^+ \p J^3 \p J^3 \nonumber\\
&&\quad -120 J^+ \p^2 J^3 J^3 +66 J^+ \p^3 J^3 +24 \p J^+ J^3 J^3 J^3 \nonumber\\
&&\quad +180 \p J^+ J^+ J^3 J^- -36 \p J^+ J^+ \p J^- -72 \p J^+ \p J^3 J^3 \nonumber\\
&&\quad +72 \p J^+ \p J^+ J^- -120 \p J^+ \p^2 J^3 -108 \p^2 J^+ J^3 J^3 \nonumber\\
&&\quad \left. +36 \p^2 J^+ J^+ J^- +36 \p^2 J^+ \p J^3 -72 \p^3 J^+ J^3+6 \p^4 J^+\right) \nonumber\\
G^{0}(z) H^{0}(w) & \sim & \f{-864 J^3 J^3}{(z-w)^4}+ \f{-864 \p J^3 J^3}{(z-w)^3} \nonumber\\
&&+ \f{144 J^3 J^3 J^3 J^3 +432 \p J^3 \p J^3 -720 \p^2 J^3 J^3 }{(z-w)^2}\\
&&+ \f{288 \p J^3 J^3 J^3 J^3+288 \p^2 J^3 \p J^3 -288 \p^3 J^3 J^3}{(z-w)} \nonumber\\
G^{0}(z) H^{-}(w) & \sim & \f{864 J^-}{(z-w)^5}+ \f{-864 J^3 J^-}{(z-w)^4} \nonumber\\
&&+ \f{-864 J^3 J^3 J^- -1728 \p J^3 J^-}{(z-w)^3}\nonumber\\
&&+ \f{288 J^3 J^3 J^3 J^- -864 \p J^3 J^3 J^- -1008 \p^2 J^3 J^- }{(z-w)^2}\nonumber\\
&&+ \f{1}{z-w} \left( 252 J^3 J^3 J^3 J^3 J^- +864 J^3 J^3 J^3 \p J^- +216 J^+ J^3 J^3 J^- J^- \right.\\
&&\quad -216 J^+ J^3 \p J^- J^- -36 J^+ J^+ J^- J^- J^-+432 J^+ \p J^3 J^- J^- \nonumber\\
&&\quad -216 J^+ \p J^- \p J^- +108 J^+ \p^2 J^- J^- -216 \p J^3 J^3 J^3 J^- \nonumber\\
&&\quad +1296 \p J^3 J^3 \p J^- -540 \p J^3 \p J^3 J^- +648 \p J^+ J^3 J^- J^- \nonumber\\
&&\quad \left. -936 \p^2 J^3 J^3 J^- +216 \p^2 J^3 \p J^- +324 \p^2 J^+ J^- J^- -468 \p^3 J^3 J^- \right) \nonumber\\
G^{0}(z) H^{--}(w) & \sim & \f{1728 J^- J^-}{(z-w)^4}+ \f{-1728 J^3 J^- J^-}{(z-w)^3}\nonumber\\
&&+ \f{-1728 J^3 \p J^- J^- -288 J^+ J^- J^- J^- -1728 \p J^- \p J^- +864 \p^2 J^- J^- }{(z-w)^2}\nonumber\\
&&+ \f{1}{z-w} \left( 576 J^3 J^3 J^3 J^- J^-+3456 J^3 J^3 \p J^- J^- +3456 J^3 \p J^- \p J^- \right.\\
&&\quad  -1728 J^3 \p^2 J^- J^- +576 J^+ J^3 J^- J^- J^- -3456 \p J^3 J^3 J^- J^- \nonumber\\
&&\quad \left. +576 \p J^+ J^- J^- J^- -1152 \p^2 J^3 J^- J^- \right) \nonumber \\
G^{-}(z) H^{++}(w) & \sim & \f{-144 J^+}{(z-w)^5}+ \f{-432 J^+ J^3-288 \p J^+}{(z-w)^4} \nonumber\\
&&+ \f{-144 J^+ J^3 J^3+72 J^+ J^+ J^- -576 J^+ \p J^3 -144 \p J^+ J^3-216 \p^2 J^+}{(z-w)^3}\nonumber\\
&&+ \f{1}{(z-w)^2} \left(48 J^+ J^3 J^3 J^3+72 J^+ J^+ J^3 J^-+72 J^+ J^+ \p J^- \right.\\
&&\quad  -288 J^+ \p J^3 J^3-384 J^+ \p^2 J^3 +144 \p J^+ J^+ J^- \nonumber\\
&& \quad \left. -144 \p^2 J^+ J^3-96 \p^3 J^+ \right) \nonumber\\
&& + \f{1}{z-w} \left(54 J^+ J^3 J^3 J^3 J^3+36 J^+ J^+ J^3 J^3 J^--36 J^+ J^+ J^3 \p J^- \right. \nonumber\\
&& \quad-18 J^+ J^+ J^+ J^- J^-+144 J^+ J^+ \p J^3 J^-+36 J^+ J^+ \p^2 J^-  \nonumber\\
&& \quad+108 J^+ \p J^3 J^3 J^3 -54 J^+ \p J^3 \p J^3 -216 J^+ \p^2 J^3 J^3 \nonumber\\
&&\quad-162 J^+ \p^3 J^3 +24 \p J^+ J^3 J^3 J^3+252 \p J^+ J^+ J^3 J^-\nonumber\\
&&\quad+36 \p J^+ J^+ \p J^- +72 \p J^+ \p J^3 J^3+72 \p J^+ \p J^+ J^-\nonumber\\
&&\quad+24 \p J^+ \p^2 J^3 -36 \p^2 J^+ J^3 J^3+180\p^2 J^+ J^+ J^-\nonumber\\
&&\quad \left. -36 \p^2 J^+ \p J^3 -72 \p^3 J^+ J^3-30 \p^4 J^+ \right) \nonumber\\
G^{-}(z) H^{+}(w) & \sim & \f{-288 J^3}{(z-w)^5}+ \f{576 J^3 J^3+144 J^+ J^--288 \p J^3}{(z-w)^4}\nonumber\\
&&+ \f{1}{(z-w)^3} \left(288 J^3 J^3 J^3-288 J^+ J^3 J^-+144 J^+ \p J^-  \right.\nonumber\\
&&\left. \quad +864 \p J^3 J^3 -288 \p J^+ J^--144 \p^2 J^3 \right)\nonumber\\
&&+ \f{1}{(z-w)^2} \left( -48 J^3 J^3 J^3 J^3-144 J^+ J^3 J^3 J^- -288 J^+ J^3 \p J^- \right. \\
&&\quad -144 J^+ \p J^3 J^-+72 J^+ \p^2 J^- +576 \p J^3 J^3 J^3 \nonumber\\
&&\quad -144 \p J^3 \p J^3 -288 \p J^+ J^3 J^--288 \p J^+ \p J^- \nonumber\\
&& \quad \left. +672 \p^2 J^3 J^3-144 \p^2 J^+ J^--48 \p^3 J^3 \right) \nonumber\\
&&+ \f{1}{z-w} \left( -84 J^3 J^3 J^3 J^3 J^3+24 J^+ J^3 J^3 J^3 J^-+288 J^+ J^3 J^3 \p J^- \right.\nonumber\\
&&\quad -144 J^+ J^3 \p^2 J^- +108 J^+ J^+ J^3 J^- J^--576 J^+ \p J^3 J^3 J^-\nonumber\\
&&\quad +72 J^+ \p J^3 \p J^- -192 J^+ \p^2 J^3 J^- +24 J^+ \p^3 J^- \nonumber\\
&& -120 \p J^3 J^3 J^3 J^3+36 \p J^3 \p J^3 J^3-360 \p J^+ J^3 J^3 J^-\nonumber\\
&&\quad+144 \p J^+ J^3 \p J^- +216 \p J^+ J^+ J^- J^--360 \p J^+ \p J^3 J^-\nonumber\\
&&\quad -144 \p J^+ \p^2 J^- +456 \p^2 J^3 J^3 J^3-120 \p^2 J^3 \p J^3 -468 \p^2 J^+ J^3 J^-\nonumber\\
&&\quad\left. -36 \p^2 J^+ \p J^-  +300 \p^3 J^3 J^3-120 \p^3 J^+ J^--12 \p^4 J^3 \right) \nonumber\\
G^{-}(z) H^{0}(w) & \sim & \f{-864 J^-}{(z-w)^5}+ \f{-864 J^3 J^--864 \p J^-}{(z-w)^4}\nonumber\\
&&+ \f{864 J^3 J^3 J^--864 J^3 \p J^- +864 \p J^3 J^--432 \p^2 J^-}{(z-w)^3}\nonumber\\
&&+ \f{1}{(z-w)^2} \left(288 J^3 J^3 J^3 J^-+864 J^3 J^3 \p J^- -432 J^3 \p^2 J^- \right.\nonumber\\
&&\quad+864 \p J^3 J^3 J^-+864 \p J^3 J^3 J^-+864 \p J^3 \p J^- +288 \p^2 J^3 J^-\nonumber\\
&&\quad\left. -144 \p^3 J^-\right)\\
&&+ \f{1}{z-w} \left(-252 J^3 J^3 J^3 J^3 J^--576 J^3 J^3 J^3 \p J^- +432 J^3 J^3 \p^2 J^- \right.\nonumber\\
&& \quad  -144 J^3 \p^3 J^- -216 J^+ J^3 J^3 J^- J^-+216 J^+ J^3 \p J^- J^-\nonumber\\
&&\quad +36 J^+ J^+ J^- J^- J^--432 J^+ \p J^3 J^- J^-+216 J^+ \p J^- \p J^- \nonumber\\
&&\quad -108 J^+ \p^2 J^- J^-+1080 \p J^3 J^3 J^3 J^--432 \p J^3 J^3 \p J^- \nonumber\\
&&\quad +540 \p J^3 \p J^3 J^-+432 \p J^3 \p^2 J^- -648 \p J^+ J^3 J^- J^-\nonumber\\
&&\quad +936 \p^2 J^3 J^3 J^-+72 \p^2 J^3 \p J^- -324 \p^2 J^+ J^- J^-+180 \p^3 J^3 J^- \nonumber\\
&&\quad  \left. -36 \p^4 J^- \right)  \nonumber\\
G^{-}(z) H^{-}(w) & \sim & \f{-2592 J^- J^-}{(z-w)^4}+ \f{-2592 \p J^- J^-}{(z-w)^3}\nonumber\\
&&+ \f{1}{(z-w)^2} \left(864 J^3 J^3 J^- J^-+1728 J^3 \p J^- J^-+288 J^+ J^- J^- J^-\right.\nonumber\\
&& \quad\left.-864 \p J^3 J^- J^-+1728 \p J^- \p J^- -2160 \p^2 J^- J^- \right) \nonumber\\
&&+ \f{1}{z-w} \left( 864 J^3 J^3 \p J^- J^-+864 J^3 \p J^- \p J^- +864 J^3 \p^2 J^- J^-\right.\\
&&\quad+432 J^+ \p J^- J^- J^-+864 \p J^3 J^3 J^- J^-+144 \p J^+ J^- J^- J^-\nonumber\\
&&\quad\left.-432 \p^2 J^3 J^- J^-+1296 \p^2 J^- \p J^- -864 \p^3 J^- J^- \right) \nonumber\\
G^{-}(z) H^{--}(w) & \sim & \f{-1728 J^- J^- J^-}{(z-w)^3}+ \f{576 J^3 J^- J^- J^--1728 \p J^- J^- J^-}{(z-w)^2}\\
&&+ \f{1}{z-w} \left( 576 J^3 J^3 J^- J^- J^-+5184 J^3 \p J^- J^- J^-+576 J^+ J^- J^- J^- J^-\right.\nonumber\\
&& \quad \left. -4608 \p J^3 J^- J^- J^-+6912 \p J^- \p J^- J^--4320 \p^2 J^- J^- J^-\right) \nonumber\\
G^{--}(z) H^{++}(w) & \sim & \f{576 J^3}{(z-w)^5}+ \f{288 J^3 J^3-576 J^+ J^-+864 \p J^3}{(z-w)^4}\nonumber\\
&&+ \f{-288 J^+ J^3 J^--864 J^+ \p J^- +576 \p J^3 J^3+576 \p^2 J^3}{(z-w)^3}\\
&&+ \f{1}{(z-w)^2} \left( -48 J^3 J^3 J^3 J^3-144 J^+ J^3 \p J^- +72 J^+ J^+ J^- J^-\right. \nonumber\\
&&\quad -432 J^+ \p J^3 J^--576 J^+ \p^2 J^- +144 \p J^3 \p J^3 -144 \p J^+ J^3 J^-\nonumber\\
&&\quad \left. +144 \p J^+ \p J^- +384 \p^2 J^3 J^3-144 \p^2 J^+ J^-+240 \p^3 J^3 \right)\nonumber\\
&&+ \f{1}{z-w} \left( -24 J^3 J^3 J^3 J^3 J^3+48 J^+ J^3 J^3 J^3 J^-+288 J^+ J^3 J^3 \p J^- \right. \nonumber\\
&&\quad +72 J^+ J^+ J^3 J^- J^-+144 J^+ J^+ \p J^- J^--288 J^+ \p J^3 J^3 J^-\\
&&\quad -144 J^+ \p J^3 \p J^- -384 J^+ \p^2 J^3 J^--240 J^+ \p^3 J^- \nonumber\\
&&\quad -144 \p J^3 J^3 J^3 J^3-72 \p J^3 \p J^3 J^3-144 \p J^+ J^3 J^3 J^-\nonumber\\
&&\quad +144 \p J^+ J^+ J^- J^--144 \p J^+ \p J^3 J^-+144 \p J^+ \p^2 J^-\nonumber\\
&&\quad +48 \p^2 J^3 J^3 J^3+144 \p^2 J^3 \p J^3 -216 \p^2 J^+ J^3 J^-\nonumber\\
&&\quad \left. -72 \p^2 J^+ \p J^-+168 \p^3 J^3 J^3-96 \p^3 J^+ J^-+72 \p^4 J^3 \right) \nonumber\\
G^{--}(z) H^{+}(w) & \sim & \f{576 J^-}{(z-w)^5}+ \f{1728 J^3 J^-+1152 \p J^-}{(z-w)^4}\nonumber\\
&&+ \f{1}{(z-w)^3} \left(576 J^3 J^3 J^-+2304 J^3 \p J^- -288 J^+ J^- J^-\right. \nonumber\\
&&\quad \left. +576 \p J^3 J^-+864 \p^2 J^- \right) \nonumber\\
&&+ \f{1}{(z-w)^2} \left( -192 J^3 J^3 J^3 J^-+1440 J^3 \p^2 J^- -288 J^+ J^3 J^- J^- \right.\nonumber\\
&&\quad -576 J^+ \p J^- J^-+1152 \p J^3 J^3 J^--288 \p J^+ J^- J^-\nonumber\\
&&\quad \left. +672 \p^2 J^3 J^-+384 \p^3 J^-\right) \\
&&+ \f{1}{z-w} \left( -216 J^3 J^3 J^3 J^3 J^--960 J^3 J^3 J^3 \p J^- -288 J^3 J^3 \p^2 J^- \right. \nonumber\\
&&\quad+576 J^3 \p^3 J^- -144 J^+ J^3 J^3 J^- J^--144 J^+ J^3 \p J^- J^-\nonumber\\
&&\quad+72 J^+ J^+ J^- J^- J^--576 J^+ \p J^3 J^- J^-+144 J^+ \p J^- \p J^- \nonumber\\
&&\quad-504 J^+ \p^2 J^- J^-+432 \p J^3 J^3 J^3 J^--288 \p J^3 J^3 \p J^- \nonumber\\
&&\quad+216 \p J^3 \p J^3 J^--288 \p J^3 \p^2 J^- -720 \p J^+ J^3 J^- J^-\nonumber\\
&&\quad-576 \p J^+ \p J^- J^-+1296 \p^2 J^3 J^3 J^-+336 \p^2 J^3 \p J^- \nonumber\\
&&\quad \left. -360 \p^2 J^+ J^- J^-+360 \p^3 J^3 J^-+120 \p^4 J^-\right) \nonumber\\
G^{--}(z) H^{0}(w) & \sim & \f{1728 J^- J^-}{(z-w)^4}+ \f{1728 J^3 J^- J^-+3456 \p J^- J^-}{(z-w)^3}\nonumber\\
&&+ \f{1}{(z-w)^2} \left(1728 J^3 \p J^- J^--288 J^+ J^- J^- J^- \right.\nonumber\\
&&\quad\left. +1728 \p J^3 J^- J^- +2592 \p^2 J^- J^- \right) \nonumber\\
&&+ \f{1}{z-w} \left( -576 J^3 J^3 J^3 J^- J^--3456 J^3 J^3 \p J^- J^-\right.\\
&&\quad-3456 J^3 \p J^- \p J^- +1728 J^3 \p^2 J^- J^--576 J^+ J^3 J^- J^- J^-\nonumber\\
&&\quad-864 J^+ \p J^- J^- J^- +3456 \p J^3 J^3 J^- J^-+1728 \p J^3 \p J^- J^-\nonumber\\
&&\quad-864 \p J^+ J^- J^- J^- +2016 \p^2 J^3 J^- J^--864 \p^2 J^- \p J^- \nonumber\\
&&\quad\left. +1440 \p^3 J^- J^-\right) \nonumber\\
G^{--}(z) H^{-}(w) & \sim & \f{1728 J^- J^- J^-}{(z-w)^3} + \f{576 J^3 J^- J^- J^-+3456 \p J^- J^- J^-}{(z-w)^2}\nonumber\\
&&+ \f{1}{z-w} \left(-576 J^3 J^3 J^- J^- J^--3456 J^3 \p J^- J^- J^-\right. \\
&& -576 J^+ J^- J^- J^- J^-+5184 \p J^3 J^- J^- J^--5184 \p J^- \p J^- J^-\nonumber\\
&& \left. +5184 \p^2 J^- J^- J^-\right) \nonumber\\
G^{--}(z) H^{--}(w) & \sim & \f{576 J^- J^- J^- J^- }{(z-w)^2}+ \f{1152 \p J^- J^- J^- J^-}{(z-w)}
\eea






\end{document}